\documentclass[12pt]{iopart}
\usepackage{amsfonts,bm}
\begin{document}
\jl{1}
\submitted
\letter{Exact results for some Madelung type
constants in the finite-size scaling theory}
\author{Hassan Chamati\footnote{e-mail: chamati@issp.bas.bg}
and Nicholay S Tonchev\footnote{e-mail: tonchev@issp.bas.bg}}
\address{Institute of Solid State Physics,
72 Tzarigradsko chauss\'ee, 1784 Sofia, Bulgaria}
\begin{abstract}
A general formula is obtained from which the Madelung type constant:
\begin{displaymath}
{\cal C}(d|\nu)=\int_0^\infty dx x^{d/2-\nu-1}
\left[\left(\sum_{l=-\infty}^\infty e^{-xl^2}\right)^d-1
-\left(\frac\pi x\right)^{\frac d2}\right]
\end{displaymath}
extensively used in the finite-size scaling theory is computed
analytically for some particular cases of the parameters $d$ and
$\nu$. By adjusting these parameters one can obtain different physical
situations corresponding to different geometries and magnitudes of the
interparticle interaction.
\end{abstract}
\pacs{05.30.-d, 05.70.Fh, 05.70.Jk, 64.60.-i}
\section*{}
In the analytic investigation of the finite-size scaling theory of
systems undergoing phase transition the Madelung type
constant~\cite{chamati96}
\begin{eqnarray}\label{madelung}
\fl \ \ \ \ \ \ \ \ \ \ \ {\cal C}(d|\nu)&=&\int_0^\infty dx x^{d/2-\nu-1}
\left[\left(\sum_{l=-\infty}^\infty e^{-xl^2}\right)^{d}-1
-\left(\frac\pi x\right)^{\frac d2}\right]\nonumber\\
&=&\lim_{\lambda\to0}\left\{{\sum_{\bm l}}'\frac{\Gamma\left(\frac
d2-\nu,\lambda{\bm l}^2\right)} {{\bm
l}^{d-2\nu}}-\int\limits_{-\infty}^\infty\cdots
\int\limits_{-\infty}^\infty d^{d}{\bm l}\frac{\Gamma\left(\frac d2-\nu,
\lambda{\bm l}^2\right)} {{\bm l}^{d-2\nu}}\right\}, \
\frac d2>\nu>0,
\end{eqnarray}
where ${\bm l}\in\mathbb Z^d$ and $\Gamma(a,x)$ is the incomplete
gamma function, plays a central role. By adjusting the parameters $d$
and $\nu$, one can obtain constants, describing different physical
situations. These situations correspond to different geometries
(hypercube, slab geometry and many others) and interparticle
interaction in the system (short as well long-range). The particular
case ${\cal C}\left(4|1\right)$, corresponding to the short-range
forces, has been widely used in the asymptotic analysis of the
finite-size properties of the ${\cal O}(n)$-symmetric $\varphi^4$
model using renormalization group treatment of
static~\cite{brezin85,zinnjustin96,chen98} as well
dynamic~\cite{goldschmidt87,niel87,Diehl87} critical phenomena. The
constant ${\cal C}\left(d|\nu\right)$ for the long-range case has been
obtained in the asymptotic analysis of finite-size effects of the
spherical model of Berlin and
Kac~\cite{chamati96,brankov88,brankov90,brankov91,chamati98} as well
as the quantum $\varphi^4$ model~\cite{chamati001} in the large $n$
limit. The same constant is obtained in the renormalization group
treatment of the finite-size scaling in ${\cal O}(n)$-symmetric
systems~\cite{chamati002}.

The constant ${\cal C}(4|1)$ is evaluated numerically with a very good
accuracy. It is found to be (see e.g. \cite{brezin85})
\begin{equation}\label{numvalue}
{\cal C}(4|1)=-1.7650848012...\ \pi=-5.545177444... .
\end{equation}

It is the aim of this letter to find a general formula for the
analytic evaluation of the constant ${\cal C}(d|\nu)$ and subsequently
to deduce useful expression in some particular cases of the parameters
$d$ and $\nu$. Let us note that the integral ${\cal C}(d|\nu)$ has the
remarkable symmetry property
\begin{equation}\label{symmetry}
\pi^\nu{\cal C}\left(d|\nu\right)=\pi^{d/2-\nu}
{\cal C}\left(d\left|\frac d2-\nu\right.\right),
\end{equation}
which relates the values of ${\cal C}\left(d|\nu\right)$ for
$\nu>\frac d4$ with those for $\nu<\frac d4$.
Equation~(\ref{symmetry}) is obtained as a consequence of using the
Jacobi identity for the sum in the integrand.

Our key finding is that the Madelung type constant can be expressed in
terms of the analytic continuation, over $\nu<\frac d2$, of
\begin{equation}\label{important}
{\cal C}\left(d|\nu\right)=\pi^{\frac d2-2\nu} \Gamma\left(\nu\right)
{\sum_{\bm l}}'\ {\bm l}^{-2\nu}, \ \ \ \ \ \ \ \ \nu>\frac d2,
\end{equation}
where ${\bm l}\in\mathbb Z^d$ and the primed summation indicates that
the term corresponding to ${\bm l}\neq0$ is excluded. For some
particular values of the dimension of the lattice, the $d$-fold sum
can be expressed as a product of simple sums such as Ditrichlet
series~\cite{zucker75}.

To show that the Madelung type constant (\ref{madelung}) is related
with the sum given in equation~(\ref{important}) we start from the
generalized $d$-dimensional Jacobi identity:
\begin{equation}\label{jacobi}
\sum_{\bm l}\exp\left(-u{\bm l}^2\right)=
\left(\frac\pi u\right)^{d/2}\sum_{\bm l}\exp\left(-\frac{\pi^2{\bm l}^2}
u\right), \ \ \ \ \ \ {\bm l}\in\mathbb Z^d.
\end{equation}
Following reference~\cite{singh89}, we multiply both sides of
(\ref{jacobi}) by $u^{d/2-\nu-1}$ and integrate over $u$. Whence, we
get the key identity (valid for $\nu\neq0,\frac d2$)
\begin{equation}\label{key}
\fl {\cal C}(d|\nu)=-\frac{u^{\frac d2-\nu}}{\frac d2-\nu}+{\sum_{\bm
l}}'\frac{\Gamma\left(\frac{d}{2}-\nu,u{\bm l}^{2}\right)}{{\bm l}
^{d-2\nu}}- \frac{\pi^\frac d2}{\nu u^\nu}
+\pi^{d/2-2\nu}{\sum_{\bm l}}'
\frac{\Gamma\left(\nu,\frac{\pi^{2}{\bm l}^2}{u}\right)}
{{\bm l}^{2\nu}}.
\end{equation}

The right hand side ${\cal C}(d|\nu)$ of identity (\ref{key}) is a
constant of integration independent of $u$. Consequently the right
hand side should also be $u$ independent. By adjusting the parameter
$u$ one obtains different expressions for the constant ${\cal
C}(d|\nu)$. All these expressions are equivalent in the sense that
they give the same `numerical' value for fixed $d$ and $\nu$. In
particular, we find it is useful to obtain simple expressions for the
constant ${\cal C}(d|\nu)$ corresponding to the limiting cases
$u\to\infty$ and $u\to0$. With the aid of the asymptotic behavior of
the incomplete gamma function~\cite{abramovitz70}
\begin{equation}\label{sol1}
\Gamma(a,x)=\left\{
\begin{array}{ll}
x^{a-1}e^{-x}\left[1-\frac{a-1}{x} +{\cal
O}\left(\frac1{x^2}\right)\right], &x\gg1,\\[.5cm]
\Gamma(a)-\frac{x^a}ae^{-x}\left[1+\frac{x}{a+1}
+{\cal O}\left(x^2\right)\right], \ \ \ &x\ll1.
\end{array}
\right.
\end{equation}
it is possible to perform the evaluation of the right hand side of
(\ref{key}) and one gets (\ref{madelung}) (valid for $0<\nu<\frac d2$)
in the limit $u\to0$, and (\ref{important}) (valid for $\nu>\frac d2$)
in the limit $u\to\infty$. It is not difficult to see that the results
(\ref{madelung}) and (\ref{important}) are valid in two different
intervals and so they complement each other.

On the other hand the sum in the right hand side of
equation~(\ref{important}) can be expressed in terms of the Epstein
zeta function~\cite{glasser80}
\begin{equation}\label{epstein}
{\cal Z} \left |\!\!\! \begin{array}{c} 0\\0
\end{array} \!\!\!\right| (d,\nu) =
{\sum_{\bm l}}'\ {\bm l}^{-2\nu}, \ \ \ \ \ \ \ {\bm l}\in\mathbb Z^d,
\ \ \ \ \ \nu>\frac d2,
\end{equation}
which can be regarded as the generalized $d$-dimensional analog of the
Riemann zeta function $\zeta(\nu)$. In the case under consideration
the Epstein zeta function has a simple pole at $\nu=\frac d2$ and may
be analytically continued in the interval $\nu<\frac d2$. Note that,
from the functional equation for the Epstein
function~\cite{glasser80}, one can check easily that ${\cal
C}\left(d|\nu\right)$, defined in~(\ref{important}), obeys the
symmetry property~(\ref{symmetry}).

Using the results of~\cite{zucker75,singh89}, for the Epstein zeta
function (\ref{epstein}), we get simple expressions, for ${\cal
C}(d|\nu)$, for certain values of $d$:

({\it a}) For the simplest one dimensional case, $d=1$, we obtain
\begin{equation}\label{d=1}
{\cal C}(1|\nu)=2\pi^{1/2-2\nu}\Gamma(\nu)\zeta(2\nu)\ \ \ \ \
\nu\neq0,\frac12.
\end{equation}
As a particular case, we give here the value of the constant ${\cal
C}(1|1/4)=2\Gamma(1/4)
\zeta(1/2)=-10.589351...$, corresponding to the short-range case with
$\nu=\frac14$.

({\it b}) In the two dimensional case, $d=2$, we obtain
\begin{equation}\label{d=2}
{\cal C}(2|\nu)=4\pi^{1-2\nu}\Gamma(\nu)\zeta(\nu)\beta(\nu)\ \ \ \ \
\nu\neq0,1,
\end{equation}
where $\beta(\nu)$ is the analytic continuation of the Dirichlet
series:
$$
\beta(\nu)=\sum_{l=0}^\infty(-1)^l(2l+1)^\nu, \ \ \ \ \nu>0.
$$
Note here that in the particular case corresponding to a long-range
potential with $\nu=\frac12$, we get ${\cal
C}(2|1/2)=4\sqrt\pi\zeta(1/2)\beta(1/2)=-6.913039577...$.

({\it c}) The constant ${\cal C}(d|\nu)$ in the four dimensional case,
$d=4$, turns out to be
\begin{equation}\label{d=4}
{\cal C}(4|\nu)=8\left(1-4^{1-\nu}\right)\pi^{2(1-\nu)}\Gamma(\nu)
\zeta(\nu-1)\zeta(\nu) \ \ \ \ \nu\neq0,2.
\end{equation}
Particularly we are interested in the value of the constant in the
case of short-range interaction corresponding to $\nu=2$. We find it
to be exactly~\cite{singh89}
$$
{\cal C}(4|1)=-8\ln2.
$$
As we mentioned above this constant has been widely used in the
analytic investigation of finite-size scaling and its relation to
numerical analysis.

Now we turn our attention to the three dimensional case. To our
knowledge there are no analytic expressions for it up to now. This
interesting case has been investigated numerically in
reference~\cite{glasser80}. In order to evaluate numerically the
constant ${\cal C}(d|\nu)$, here we propose the following general
formula
\begin{equation}\label{for}
\fl {\cal C}(d|\nu)=-\frac{u^{\frac d2-\nu}}{\frac d2-\nu}+{\sum_{\bm
l}}'\frac{\Gamma\left(\frac{d}{2}-\nu,u{\bm l}^{2}\right)}{{\bm l}
^{d-2\nu}}- \frac{\pi^\frac d2}{\nu u^\nu}
+{\cal O}\left(u^{\nu-1}\exp\left[-\frac{\pi^2}{u}\right]\right)
\end{equation}
obtained from equation (\ref{key}) for $u$ small enough. Equation
(\ref{for}) is a generalization of a three dimensional result obtained
in reference \cite{chaba75}. Note that equation (\ref{for}) is valid
for arbitrary $d$ and $\nu\neq0,\frac d2$, which makes it suitable for
numerical evaluations of the constant ${\cal C}(d|\nu)$, especially in
the cases of 3d and 5d used in finite-size scaling.

The nature of the error term equation (\ref{for}) is such that this
formula is useful for numerical evaluations even when the parameter
$u$ is not too small. Since ${\cal C}(d|\nu)$ with concrete values of
the parameters $d$ and $\nu$ are related with finite-size properties
of confined systems with different geometries and different types of
interparticle interaction, we shall present some most useful values
obtained numerically from equation (\ref{for}).

In the case of short-range interaction two constants were used in the
literature corresponding to different situations. For a three
dimensional system confined to a fully finite geometry we find the
value ${\cal C}(3|1/2)=-8.91362917...$ in perfect agreement with
references \cite{chen98,brankov88,chaba75,monkhrost70}. For a system
with a slab geometry, we get ${\cal C}(3|1)=-5.028978843...$,
coinciding with the value of reference \cite{brezin85}. Another case
of interest is that of system with long-range case $\nu=3/4$ and a
cubic geometry. For this case we have ${\cal
C}(3|3/4)=-5.9098415587...$. At the end, we quote a result for the
$5d$ case, it is ${\cal C}(5|1)=-4.228709895...$ in agreement with
that of reference~\cite{chen98}, where this constant was used in the
investigation of a the finite scaling scaling in a five dimensional
system confined to a cubic geometry.

Numerical values of ${\cal C}(4\nu|\nu)/
\Gamma(\nu)$ for $\nu=\frac14,\ \frac12,\ \frac34$ and the exact
value for $\nu=1$ were recently reported in reference
\cite{luijten99}.

This work is supported by The Bulgarian Science Foundation under grant
F608/96.

\section*{Note added in proof}
After the completion of this paper we learned that in a particular
form, identity~\ref{important} was first used by Epstein in 1903. We
are thankful to R.M. Ziff, who brings to our attention this fact and
the review article on ideal Bose-Einstein gas in
reference~\cite{ziff77}, where the same type of functions appear in
the case of finite volume.

\end{document}